\documentclass[12pt]{JHEP3}
\usepackage{mathrsfs}
\usepackage{amsmath,amssymb}
\usepackage{epsfig}
\usepackage{relsize}
\usepackage{graphicx}
\input epsf

\usepackage{epsfig}
\usepackage{relsize}

\input epsf

\usepackage{mathrsfs}
\usepackage{amsmath,amssymb}
\usepackage{epsfig}
\input epsf

\def\x'{\mathaccent 19 x}
\def\y'{\mathaccent 19 y}
\def\n'{\mathaccent 19 n}
\def\u'{\mathaccent 19 u}

\def\et'{\mathaccent 19 \eta}
\def\th'{\mathaccent 19 \theta}
\def\lam'{\mathaccent 19 \lambda}
\def\varet'{\mathaccent 19 \vartheta}
\def\rh'{\mathaccent 19 \rho}
\def\ph'{\mathaccent 19 \phi}
\def\xb'{\mathaccent 19 {\bar{x}}}

   \def \tr {{\rm tr}}

\def \A {{\cal{A}}}

\def\det{\hbox{det}}
\def\be{\begin{equation}}
\def\ee{\end{equation}}

\newcommand{\bea}{\begin{eqnarray}}
\newcommand{\eea}{\end{eqnarray}}

\def\r {\rho}
\def\a {\alpha}
\def\b {\beta}
\def\s {\sigma}
\def\pa {\partial}

\def\k{\kappa}

\newcommand{\alg}[1]{\mathfrak{#1}}
\newcommand{\su}{\alg{su}}

\newcommand{\AdS}{{\rm  AdS}_5\times {\rm S}^5}

\def\L{\mathscr L}

\newcommand{\sfrac}[2]{{\textstyle\frac{#1}{#2}}}

\def \B {\text B}

\def\ov{\over}

\def\la{\label}

\def\CP{\mathbb{ CP}}

\def\osp{\alg{osp}(2,2|6)}

\newcommand{\ads}{{\rm  AdS}}

\author{Gleb Arutyunov$^a$\footnote{Email: G.Arutyunov@phys.uu.nl, frolovs@maths.tcd.ie} {}
\footnote{Correspondent fellow at Steklov Mathematical Institute,
Moscow.} \  and Sergey Frolov$^{b\, \dagger}$
 \\ $^{a}$ {\it Institute for Theoretical
Physics and Spinoza Institute,\\ ~~Utrecht University, 3508 TD
Utrecht, The Netherlands} \\ $^b$ {\it School of Mathematics,
Trinity College, Dublin 2, Ireland}}

\abstract{According to the recent proposal by Aharony, Bergman,
Jafferis and Maldacena the ${\cal N}=6$ supersymmetric
Chern-Simons theory in three dimensions has a 't Hooft limit whose
holographic dual is described by type IIA superstings on ${\rm
AdS}_4\times \CP^3$ background. We argue that the Green-Schwarz
action for type IIA string theory  on ${\rm AdS}_4\times \CP^3$
with $\kappa$-symmetry {\it partially fixed} can be understood as
a coset sigma-model on the same space
supplied with a proper Wess-Zumino term. We construct the
corresponding sigma-model Lagrangian and show that it is invariant
under a local fermionic symmetry which for generic bosonic string
configurations allows one to remove 8 out of 24 fermionic degrees of
freedom. The remaining 16 fermions together with their bosonic
partners should describe the physical content of ${\rm
AdS}_4\times \CP^3$ superstring. As further evidence, we
demonstrate that in the plane-wave limit the quadratic action
arising from our model reproduces the one emerging from the type
IIA superstring. The coset sigma-model is classically integrable
which opens up the possibility to investigate its dynamics in a way
very similar to the case of $\AdS$ superstrings. }

\title{
Superstrings on {\bf AdS}$_4\times {\bf CP}^3 $ as \\
a Coset Sigma-model }  \preprint{
          \smaller{\smaller{\smaller{ITP-UU-08-40}}}\\[-.5ex]
          \smaller{\smaller{\smaller{SPIN-08-31}}}\\[-.5ex]
          \smaller{\smaller{\smaller{TCDMATH 08-08}}}}

\begin{document}

\section{Introduction and Summary}
Superconformal Chern-Simons theories \cite{Schwarz:2004yj}
conjectured to describe the low-energy world-volume dynamics of
multiple M2-branes are receiving nowadays considerable attention
\cite{Bagger:2006sk}. Recently Aharony, Bergman, Jafferis and
Maldacena (ABJM) proposed a new example of the AdS/CFT duality
which involves the ${\cal N}=6$ superconformal ${\rm SU}(N)\times
{\rm SU}(N)$ Chern-Simons theory in three dimensions and the
M-theory on ${\rm AdS}_4\times {\rm S}^7/{\mathbb Z}_k$, where $k$
is the level of the Chern-Simons action \cite{Aharony:2008ug}.

\smallskip

The ABJM model is characterized by two parameters -- the rank $N$
of the two gauge groups ${\rm SU}(N)$ and the integer level $k$
which is opposite for  the  gauge groups. Remarkably, there exists
an analogue of the 't Hooft limit, where $N,k\to\infty$ with the
ratio $\lambda=2\pi^2 N/k$ kept fixed. In this limit $\lambda$
becomes continuous allowing therefore for application of standard
perturbative techniques. It turns out that at leading order in the
weak coupling expansion the corresponding dilatation operator can
be identified with an integrable Hamiltonian of the ${\rm SU }(4)$
spin chain with spins alternating between fundamental and
anti-fundamental representations \cite{Minahan:2008hf}. The set of
emerging Bethe equations admits an extension to the full
superconformal group ${\rm OSP}(2,2|6)$.

\smallskip

According to \cite{Aharony:2008ug}, at strong coupling, {\it i.e.}
when $\lambda$ becomes large, the M-theory on ${\rm AdS}_4\times
{\rm S}^7/{\mathbb Z}_k$ can be effectively described by type IIA
superstring theory on the ${\rm AdS}_4\times \CP^3$ background. To
understand this new example of holography, one needs, therefore,
to determine the spectrum of the corresponding string theory.
Obviously, the classical bosonic string theory can be  formulated
as a sigma-model on the ${\rm AdS}_4\times \CP^3$ space. This
model is integrable but quantum corrections related to  $\CP^3$
are known to spoil its classical integrability
\cite{Abdalla}. One may hope,
however, that inclusion of type IIA fermions in the full model
would maintain integrability at the quantum level. This is not an
easy question to answer. The presence of the background RR fields
sustaining the ${\rm AdS}_4\times \CP^3$ metric \cite{adscp3}  suggests to use
the Green-Schwarz formulation for type IIA superstrings. On the
other hand, the complete Green-Schwarz action on  ${\rm
AdS}_4\times \CP^3$ to all orders in fermionic variables  is
unknown\footnote{In principle, one can obtain the corresponding
string action by performing the double dimensional reduction of
the supermembrane action on ${\rm AdS}_4\times {\rm S}^7$
constructed in \cite{deWit:1998yu} .}. Even the knowledge of the
action alone would be of little use to expose integrable
properties of the corresponding model.

\smallskip

In this paper we propose a novel way to investigate the dynamics
of type IIA strings on ${\rm AdS}_4\times \CP^3$ and, in
particular, to reveal its classical integrability. The main idea
is to follow  closely the case of type IIB superstrings on $\AdS$,
where the coset sigma-model formulation \cite{Metsaev:1998it}
provides an alternative to the conventional Green-Schwarz
approach. The ${\rm AdS}_4\times \CP^3$ space is a coset ${\rm
SO}(3,2)/{\rm SO(3,1)\times {\rm SO}(6)/{\rm U}(3)}$. The group
${\rm SO}(3,2)\times {\rm SO}(6)$ is a bosonic subgroup of the
superconformal group ${\rm OSP}(2,2|6)$ which naturally suggests
to include fermionic degrees of freedom by considering a
sigma-model on the coset space ${\rm OSP}(2,2|6)/({\rm
SO}(3,1)\times {\rm U}(3))$.

\smallskip

A problem one immediately faces with this formulation is that the
corresponding coset space contains 24 real fermions, which is too
little in comparison to 32 fermions of the Green-Schwarz type IIA
superstring. On the other hand, because of $\kappa$-symmetry only
half of fermions are physical in the latter case. Thus, the
sigma-model we propose could be just a partially $\kappa$-symmetry
fixed version of the Green-Schwarz type IIA superstring, where
only 8 out of 32 fermions have been gauged away. To justify this
interpretation, the sigma-model in question when supplied with a
proper Wess-Zumino term must allow for a local fermionic symmetry
which removes another 8 unphysical fermions. Construction of
$\kappa$-symmetry transformations  which precisely do this job
is one of the  results of our paper.

\smallskip

We find that for generic bosonic configurations, i.e. the ones,
which involve string motion in both ${\rm AdS}_4$ and $\CP^3$
directions, the rank of $\kappa$-symmetry variations is 8. There
are, however, ``singular'' configurations corresponding to string
moving in the AdS part of the coset only. For these configurations
the rank of $\kappa$-symmetry variations is 12. We argue that the
singular nature of these string backgrounds is due to their
incompatibility with the $\kappa$-symmetry gauge choice that has to be made
in order to reduce the full-fledged Green-Schwarz superstring to
our coset model.

\smallskip To get more evidence to our interpretation, we further
derive the quadratic fermionic action arising in the expansion of
the full sigma-model action around the point-particle geodesics
and show that it precisely coincides with the one which emerges
from Penrose limit of type IIA superstings on ${\rm AdS}_4\times
\CP^3$ \cite{Sugiyama:2002tf}-\cite{Gaiotto:2008cg}.

\smallskip

It should be noted that our sigma-model construction is very close
to that for $\AdS$ superstrings. This allows us to conclude
straightforwardly  on classical integrability of the model by
exhibiting the same type of the Lax connection as was found for
$\AdS$ superstring \cite{Bena:2003wd}. We also verify that this
Lax connection is compatible with $\kappa$-symmetry we found. This
opens up the possibility to investigate (partially $\kappa$-gauge
fixed) strings on ${\rm AdS}_4\times \CP^3$ by employing the
methods built up for the $\AdS$ case. In particular, imposing the
uniform light-cone gauge \cite{Arutyunov:2004yx, FPZ} one can
develop semiclassical quantization to verify whether classical
integrability is preserved by leading quantum corrections. Then,
the subalgebra $\su(2|2)$ of the global symmetry algebra ${\rm
OSP}(2,2|6)$ which leaves the light-cone Hamiltonian invariant in
the limit of infinite light-cone momentum undergoes a central
extension by the generator $P$ of the world-sheet momentum
\cite{Beisert:2005tm,Arutyunov:2006ak}. Assuming integrability of
the  quantum sigma-model, it would be interesting to see to which
extend this symmetry of the light-cone Hamiltonian can be used to
fix the form of the scattering matrix. Finite-gap solutions
(including fermionic excitations) \cite{Kazakov:2004qf} could be
also investigated with the goal of reconstructing the data of the
string S-matrix undetermined by symmetries, e.g. the dressing
phase \cite{Arutyunov:2004vx}. In the $\AdS$ case the dressing
phase satisfies the crossing symmetry equation \cite{Janik}, which
essentially determines its form \cite{BHLBES}. It would be also
interesting to understand constraints on the $\ads_4\times\CP^3$
scattering matrix  imposed by crossing symmetry.

\smallskip

The paper is organized as follows. In the next section after a
brief introduction to the Lie algebra $\osp$ we present the
Lagrangian and equations of motion of the coset model. In section
3 we deduce the corresponding $\kappa$-symmetry transformations
and analyse the rank of on-shell $\kappa$-symmetry
transformations. In section 4 we exhibit the Lax connection for
our model and demonstrate that under $\kappa$-variations it
retains  on-shell zero curvature. Section 5 is devoted to analysis
of the quadratic action for fermions around a  null geodesics. Some
technical details are relegated to appendices A and B.

\newpage

\section{Sigma-model Lagrangian}
\subsection{Coset model and its relation to IIA superstrings}
To describe superstrings propagating in the ${\rm AdS}_4\times
\CP^3$ background, one may try to develop the corresponding
Green-Schwarz formalism \cite{Green:1983wt}. We recall that the
Green-Schwarz superstring involves two Majorana-Weyl fermions of
different chirality with the total number of 32 fermionic degrees
of freedom. On the other hand, the Green-Schwarz string action
exhibits a local fermionic symmetry ($\kappa$-symmetry), which
allows one to remove a half of them. The remaining 16 fermions are
physical and in the light-cone gauge they match  with 8 bosons
rendering the space-time supersymmetry manifest. Unfortunately,
the explicit form of the type IIA Green-Schwarz action in an
arbitrary background is known up to quartic terms only
\cite{Cvetic:1999zs} and, for this reason, it remains  unknown for
the ${\rm AdS}_4\times \CP^3$ background. Even if such an action
would be found, it would not be straightforward to reveal its
integrable properties. A great advantage of type IIB string theory
on $\AdS$  is that it admits an alternative description as a coset
sigma-model \cite{Metsaev:1998it} which allows one, in particular,
to prove its classical integrability \cite{Bena:2003wd}.

\smallskip

To make a progress in understanding the string dynamics in the
${\rm AdS}_4\times \CP^3$ space-time, we devise an approach which
does not rely on the knowledge of the Green-Schwarz action. Our
starting point is to introduce a sigma-model on the coset space
\bea\frac{{\rm OSP(2,2|6)}}{{\rm SO(3,1)}\times {\rm U(3)}}\,
.\label{sAdS}\eea
 Recall that the supergroup ${\rm OSP}(2,2|6)$ has a
bosonic subgroup ${\rm USP(2,2)}\times {\rm SO(6)}$; the quotient
of the latter over ${\rm SO(3,1)}\times {\rm U(3)}$ provides a
model of the ${\rm AdS}_4\times \CP^3$ superspace with ${\rm
SO(3,1)}\times {\rm U(3)}$ playing the role of the local Lorentz
group. The superspace obtained in this way is parametrized by 24
(real) fermion degrees of freedom, which is apparently different
from both 32 (the gauge unfixed Green-Schwarz superstring) and 16
(the gauge-fixed Green-Schwarz superstring). We then show  that
the standard kinetic term can be supplemented with the Wess-Zumino
term so that the whole action does admit a local fermionic
symmetry much analogous to the usual $\kappa$-symmetry of the
Green-Schwarz superstring. We will show that for  generic bosonic
configurations the  $\kappa$-symmetry of the coset model allows
one to gauge away precisely 8 fermions, so that the resulting
fermionic content match to that of a $\kappa$-symmetry fixed
version of the Green-Schwarz superstring. This suggests an
interpretation of the coset sigma-model with 24 fermions as {\it
the Green-Schwarz superstring with a  partial fixing of
$\kappa$-symmetry} which consists in removing 8 from 32 fermions.

\smallskip

The construction of the Lagrangian for the sigma-model in question
is very similar to that for classical superstrings on $\AdS$ space
\cite{Metsaev:1998it,Roiban:2000yy,Das:2004hy,Alday:2005gi}
 and it makes use of the ${\mathbb Z}_4$-grading of the
$\osp$ Lie algebra. We start with recalling the necessary facts
about $\osp$.

\subsection{Superalgebra $\osp$ and ${\mathbb Z}_4$-grading}
The Lie algebra $\osp$ can be realized by $10\times 10$
supermatrices of the form \bea
A=\left(\begin{array}{ll} X ~&~ \theta \\
\eta ~&~ Y   \end{array}\right) \, ,\eea where $X$ and $Y$ are
even (bosonic) $4\times 4$ and $6\times 6$ matrices, respectively.
The $4\times 6$ matrix $\theta$ and the $6\times 4$ matrix $\eta$
are odd, i.e. linear in fermionic variables. The matrix $A$ must
satisfy the following two conditions \bea\la{usp} &&
A^{st}\left(\begin{array}{cc} C_{4} ~~~&~~~ 0 \\ 0 ~~~&~~~
{\mathbb I}_{6\times 6}
\end{array}\right) + \left(\begin{array}{cc} C_{4} ~~~&~~~ 0 \\ 0
~&~ {\mathbb I}_{6\times 6} \end{array}\right)A=0 ~~~ \Rightarrow
~~~ A^{st} = - \check C A \check C^{-1}\, ,\\ \la{su}
 &&
A^{\dagger}\left(\begin{array}{cc} \Gamma^0 ~&~ 0 \\ 0 ~&~
-{\mathbb I}_{6\times 6}
\end{array}\right) + \left(\begin{array}{cc} \Gamma^0 ~&~ 0 \\ 0
~&~ -{\mathbb I}_{6\times 6} \end{array}\right)A=0 ~~~ \Rightarrow
~~~ A^{\dagger} = - \check \Gamma A \check \Gamma^{-1}\, .\eea
Here $C_4$ is the charge conjugation matrix, and $A^{st}$ denotes
the super-transpose matrix \bea
A^{st}=\left(\begin{array}{rr} X^t ~&~-\eta^t \\
 \theta^t ~&~ Y^t   \end{array}\right) \, .\eea
We have also introduced four gamma-matrices $\Gamma^{\mu}$ which
satisfy the Clifford algebra of $\alg{so}(3,1)$; their explicit
form is given in appendix A. Condition (\ref{usp}) singles out
$\alg{osp}(4|6)$ with the  bosonic subalgebra $\alg{sp}(4,{\mathbb
C})\oplus \alg{so}(6,{\mathbb C})$. Eq.(\ref{su}) defines a real
section of $\alg{osp}(4|6)$ which we denote by $\osp$.

\smallskip

The charge conjugation matrix can be chosen to be real,
skew-symmetric and satisfying $C_4^2=-\mathbb{I}$, see appendix A
for an explicit representation. Conditions (\ref{usp}) and
(\ref{su}) imply that the matrices $X$ and $Y$ have the following
transposition and reality properties \bea\label{rel1} X^t&=&-C_4 X
C_4^{-1}\, , ~~~~~~~X^*=(i\Gamma^3) X (i \Gamma^3)^{-1}\,
~~~~~~~~~i\Gamma^3=\Gamma^0C_4\,  ,\\
\label{rel2} Y^t&=&-Y\, , ~~~~~~~~~~~~~~~~~Y^*=Y\, , \eea while
$\eta$ and $\theta$ obey \bea \eta=-\theta^t C_4\, ,
~~~~~~~~~\theta^*=i\Gamma^3 \theta\, . \label{rel3}\eea

\smallskip

The algebra $\alg{osp}(4|6)$ does not admit an outer automorphism
of order four \cite{serganova}. Thus, we should search for an
inner automorphism of order four such that its stationary point
would coincide with the subalgebra $\alg{so}(3,1)\times
\alg{u}(3)$.

\smallskip

Introduce two $4\times 4$ and $6\times 6$ matrices $K_4$ and
$K_6$, respectively. We require that $K_4^2=-{\mathbb I}$ and
$K_6^2=-{\mathbb I}$. In addition, we require
$(\Gamma^{\mu})^t=K_4\Gamma^{\mu}K_4^{-1}$ for all gamma-matrices.
In what follows it is convenient to make the following choice \bea
K_4=-\Gamma^1\Gamma^2={\small \left(\begin{array}{rrrrr} 0 & 1 & 0
& 0 & \\
-1 & 0 & 0 & 0 \\
0  & 0 & 0 & 1  \\
0  & 0 & -1& 0 \\
\end{array}\right)}\, , ~~~~~~~ K_6={\small \left(\begin{array}{rrrrrrr} 0
& 1 & 0 & 0 &
0 & 0 \\
-1 & 0 & 0 & 0 & 0 & 0  \\
0  & 0 & 0 & 1 & 0 & 0  \\
0  & 0 & -1& 0 & 0 & 0 \\
0  & 0 & 0 & 0 & 0 & 1 \\
0  & 0 & 0 & 0 & -1& 0
\end{array}\right)}\, .\eea
Define a map \bea
\Omega(A)=\left(\begin{array}{rr} K_4X^tK_4 ~&~ K_4\eta^tK_6 \\
-K_6\theta^tK_4 ~&~ K_6Y^tK_6\end{array}\right)\, . \eea For any
two supermatrices $A$ and $B$ it satisfies the following property
$$
\Omega(AB)=-\Omega(B)\Omega(A)
$$
and, for this reason, it is an automorphism of $\alg{osp}(4|6)$,
i.e.
$$
\Omega([A,B])=-[\Omega(B),\Omega(A)]=[\Omega(A),\Omega(B)]\, .
$$
This automorphism is inner. Indeed, using the relations
(\ref{rel1})-(\ref{rel3}), we find that \bea \Omega(A)&=&
\left(\begin{array}{rr} K_4C_4 ~&~  0\\
0 ~&~ -K_6\end{array}\right)\left(\begin{array}{ll} X ~&~ \theta \\
\eta ~&~ Y   \end{array}\right)\left(\begin{array}{rr} K_4C_4 ~&~  0\\
0 ~&~ -K_6\end{array}\right)^{-1}\equiv \Upsilon A \Upsilon^{-1}\,
.\eea Since $(K_4C_4)^2={\mathbb I}$, and $K_6^2=-{\mathbb I}$,
the element $\Upsilon\in {\rm OSP(4|6)}$ obeys
$\Upsilon^4={\mathbb I}$. In fact, the matrix $K_4C_4$ coincides
with $\Gamma^5$ given by
$\Gamma^5=-i\Gamma^0\Gamma^1\Gamma^2\Gamma^3$. Further, we note
that $\Upsilon^{\dagger}\check \Gamma \Upsilon \check
\Gamma^{-1}={\rm diag}(-{\mathbb I}_4,{\mathbb I}_6)$. This means
that $\Omega$ does not preserve the real form $\osp$.

\medskip

The automorphism $\Omega$ allows one to endow $\A=\osp$ with the
structure of a $\mathbb{Z}_4$-graded algebra, i.e.,  as the
vector space $\A$ can be decomposed into a direct sum of four
subspaces \bea \A=\A^{(0)}\oplus \A^{(1)}\oplus \A^{(2)}\oplus
\A^{(3)}\, \label{Z4}\eea such that $[\A^{(k)},\A^{(m)}]\subseteq
\A^{(k+m)}$ modulo $\mathbb{Z}_4$. Each subspace $\A^{(k)}$  in
eq.(\ref{Z4}) is an eigenspace of $\Omega$ \bea
\Omega(\A^{(k)})=i^{k}\A^{(k)} \, . \eea  Explicitly, the
projection $A^{(k)}$ of a generic element $A\in \osp$ on the
subspace $\A^{(k)}$ is constructed as follows \bea
A^{(k)}=\frac{1}{4}\Big(A+i^{3k}\Omega(A)+i^{2k}\Omega^2(A)+i^k\Omega^3(A)\Big)\,
. \label{Z4proj} \eea In particular, the stationary subalgebra of
$\Omega$ is determined by the conditions \bea [\Gamma^5,X]=0\, ,
~~~~[K_6,Y]=0 \eea and it coincides with $\alg{so}(3,1)\times
\alg{u}(3)$, see appendix A for details.

\smallskip

The space $A^{(2)}$ is spanned by matrices satisfying the
following condition \bea \Omega(A)=\Upsilon A \Upsilon^{-1}=-A\, .
\label{adscp3}\eea As is shown in appendix A, any such matrix
satisfies the following remarkable identity
\bea \label{rA}
A^3=\sfrac{1}{8}\, {\rm str}(\Sigma A^2)\, A + \sfrac{1}{8}\, {\rm
str}( A^2)\, \Sigma A\, ,
\eea
or, equivalently,
\bea\label{identity} A^3=\sfrac{1}{8}({\tr}A_{\rm AdS}^2+{\rm
tr}A^2_{\CP})\, A+\sfrac{1}{8}({\tr}A_{\rm AdS}^2-{\rm
tr}A^2_{\CP})\, \Sigma A\, . \eea
 Here $\Sigma$ is a diagonal matrix
$\Sigma=\Upsilon^2=({\mathbb I}_4,-{\mathbb I}_6)$. Equation
(\ref{adscp3}) boils down to \bea \label{ac}\{X,\Gamma^5\}=0\, ,
~~~~\{Y,K_6\}=0 \, . \eea The first equation can be solved as
$$
X=x_{\mu}\Gamma^{\mu}\,
$$
and it provides a parametrization of the coset space ${\rm
AdS}_4={\rm SO(3,2)}/{\rm SO}(3,1)$ in terms of four unconstrained
variables $x_{\mu}$. Analogously, a general solution to the second
equation in (\ref{ac}) gives a parametrization of  $\CP^3$
$$
Y=y_iT_i,
$$
where $y_i$, $i=1,\ldots, 6$ are six unconstrained variables and
the matrices $T_i$ are described in appendix A.

\smallskip

Finally, for the reader's convenience we present an explicit form
of the projections $A^{(1)}$ and $A^{(3)}$ of the matrix $A$: \bea
A^{(1)}=\frac{1}{2}\left(\begin{array}{cc} 0 &
\theta-i\Gamma^5\theta K_6 \\
\eta+i K_6 \eta\, \Gamma^5 & 0 \end{array}\right) \,
,~~~~~A^{(3)}=\frac{1}{2}\left(\begin{array}{cc} 0 &
\theta+i\Gamma^5\theta K_6 \\
\eta-i K_6 \eta\, \Gamma^5 & 0 \end{array}\right) \, .
\nonumber\eea Each of these matrices $A^{(1)}$ and $A^{(3)}$
depend on 12 real fermionic variables.

\subsection{The Lagrangian}
Let $g$ be an element of the coset (\ref{sAdS}) realized as an
embedding in the supergroup ${\rm OSP}(2,2|6)$. We use $g$ to
build the following current (the one-form) \bea \label{la}
A=-g^{-1}{\rm d}g=A^{(0)}+A^{(2)}+A^{(1)}+A^{(3)}\, . \eea The
current takes values in the algebra $\osp$ and on the r.h.s. of the
last formula we exhibited its $\mathbb{Z}_4$-decomposition. By
construction $A$ has vanishing curvature: \bea \label{zeroc}
\pa_{\a} A_{\b}-\pa_{\b} A_{\a}-[A_{\a},A_{\b}]=0\, . \eea The
sigma-model we are looking for is then described by the following
action
 \bea \la{stac}
 S=-\frac{R^2}{4\pi \a'}\int {\rm d}\sigma{\rm d}\tau\,
\L \, ,\eea where $R$ is the radius of the AdS space and the
Lagrangian density is the sum of the kinetic and the Wess-Zumino
terms \bea \label{sLag} \L =\gamma^{\a\b}{\rm
str}\big(A^{(2)}_{\a}A^{(2)}_{\b}\big)+\kappa
\epsilon^{\a\beta}{\rm str}\big(A^{(1)}_{\a}A^{(3)}_{\beta}\big)\,
 .\eea Here we use the convention $\epsilon^{\tau\sigma}=1$ and
$\gamma^{\a\b}= h^{\a\b} \sqrt {-h}$ is the Weyl-invariant
combination of the world-sheet metric $h_{\a\beta}$ with
$\det\gamma=-1$. The parameter $\kappa$ in front of the
Wess-Zumino term is kept arbitrary for the moment. As we will see
shortly, the requirement of $\kappa$-symmetry leaves two
possibilities $\kappa=\pm 1$. We note that the invariant form
defined by means of the supertrace is non-degenerate for the
orthosymplectic groups ${\rm OSP}(2n|2n+2)$.

\smallskip

Equations of motion derived from this Lagrangian read as \bea
\label{seom} \pa_{\a}{\rm \Lambda}^{\a}-[A_{\a},\Lambda^{\a}]=0\,
, \eea where we have introduced the combination \bea
\Lambda^{\a}=\gamma^{\a\b}A^{(2)}_{\b} -\sfrac{1}{2}\kappa\,
\epsilon^{\a\beta}(A^{(1)}_{\beta}-A^{(3)}_{\beta}) \, .
\label{cS} \eea The equations of motion imply the conservation of
the Noether current $J^{\a}=g \Lambda^{\a} g^{-1}$ corresponding
to the global ${\rm OSP}(2,2|6)$ symmetry of the model:
$\pa_{\a}J^{\a}=0$.

\smallskip

Single equation (\ref{seom}) can be projected on various
eigenspaces of the $\mathbb{Z}_4$ automorphism. The projection on
the subspace $\A^{(0)}$ vanishes. For the projection on $\A^{(2)}$
we get \bea \label{Eqb} \pa_{\a}(\gamma^{\a\b}A_{\b}^{(2)})
-\gamma^{\a\b}[A_{\a}^{(0)},A_{\b}^{(2)}] +\sfrac{1}{2}\kappa
\epsilon^{\a\beta}\Big([A_{\a}^{(1)},A_{\b}^{(1)}]-[A_{\a}^{(3)},A_{\b}^{(3)}]\Big)=0\,
, \eea while for projections on $\A^{(1,3)}$ one finds
\begin{equation}
\label{Eqf1} \begin{aligned}
 & \gamma^{\a\b}[A_{\a}^{(3)},A_{\b}^{(2)}] +\kappa
\epsilon^{\a\beta}[A_{\a}^{(2)},A_{\b}^{(3)}]=0\, , \\
& \gamma^{\a\b}[A_{\a}^{(1)},A_{\b}^{(2)}] -\kappa
\epsilon^{\a\beta}[A_{\a}^{(2)},A_{\b}^{(1)}]=0\,
.\end{aligned}\end{equation} In deriving these equations we  also
used the condition of zero curvature for the connection $A_{\a}$.
Introducing the tensors \bea {\rm P}_{\pm}^{\a\b}=\sfrac{1}{2}(
\gamma^{\a\b}\pm \kappa \epsilon^{\a\b} )\, ,\eea equations
(\ref{Eqf1})  can be concisely written as
\begin{equation}
\label{Eqf} \begin{aligned}
&{\rm P}_{-}^{\a\b}[A_{\a}^{(2)},A_{\b}^{(3)}] = 0 \, ,\\
&{\rm P}_{+}^{\a\b}[A_{\a}^{(2)},A_{\b}^{(1)}] = 0 \, .
\end{aligned}\end{equation}
These are equations of motion for fermions. Note that for
$\kappa=\pm 1$ the tensors ${\rm P}_{\pm}$ are orthogonal
projectors: \bea {\rm P}_{+}^{\a\b}+{\rm
P}_{-}^{\a\b}=\gamma^{\a\b}\, ,~~~~~{\rm P}_{\pm}^{\a\delta}{\rm
P}_{\pm \delta}^{~~\b}={\rm P}_{\pm}^{\a\b} \, ,~~~~~~\, {\rm
P}_{\pm}^{\a\delta}{\rm P}_{\mp \delta}^{~~\b}=0\, .\eea

\smallskip

Finally, we also have equations of motion for the world-sheet
metric which are the Virasoro constraints: \bea \label{Vir} {\rm
str}(A^{(2)}_{\a}A^{(2)}_{\b})
-\sfrac{1}{2}\gamma_{\a\b}\gamma^{\rho\delta}{\rm
str}(A^{(2)}_{\rho}A^{(2)}_{\delta})=0\, .
 \eea
We stress that so far the construction of the coset sigma-model
does not differ from that for the $\AdS$ superstring
\cite{Metsaev:1998it,Roiban:2000yy,Das:2004hy,Alday:2005gi}. The
real problem, however, is to show that the above action enjoys a
local fermionic symmetry which is capable of gauging away
precisely eight fermionic degrees of freedom. This will be the
subject of the next section.

\section{Local Fermionic Symmetry}\la{kappasym}

\subsection{Deriving $\kappa$-symmetry} Kappa-symmetry is a local
fermionic symmetry of the Green-Schwarz superstring
\cite{Green:1983wt}. It generalizes the local fermionic symmetries
first discovered for massive and massless superparticles
\cite{deAzcarraga:1982dw,Siegel:1983hh} and its presence is
crucial to ensure the space-time supersymmetry of the physical
spectrum.  In this section we establish $\kappa$-symmetry
transformations associated with the Lagrangian (\ref{sLag}).

\smallskip

The action of the global symmetry group ${\rm OSP}(2,2|6)$ is
realized on a coset element by multiplication from the left. In
this respect, $\kappa$-symmetry transformations can be understood
as the {\it right local} action of a fermionic element $G=\exp
\epsilon$ from ${\rm OSP}(2,2|6)$ on a coset representative $g$
\cite{mc}: \bea \label{ks} gG(\epsilon)=g'g_c\, , \eea where
$\epsilon\equiv \epsilon(\sigma)$ is a local fermionic parameter.
Here $g_c$ is a compensating element from ${\rm SO(3,1)}\times
{\rm U(3)}$. The fundamental difference with the case of global
symmetry is that for arbitrary $\epsilon$ the action is not
invariant under the transformation (\ref{ks}). Below we find the
conditions on $\epsilon$ which guarantee the invariance of the
action.

\smallskip First we note that under the local multiplication from
the right the connection $A$ transforms as follows \bea
\delta_{\epsilon} A=-{\rm d}\epsilon +[A,\epsilon]\, . \eea The
$\mathbb{Z}_4$-decomposition of this equation gives \bea \nonumber
\delta_{\epsilon} A^{(1)}&=&-{\rm d}\epsilon^{(1)}
+[A^{(0)},\epsilon^{(1)}]+[A^{(2)},\epsilon^{(3)}]\,
,\\\label{dcst}\delta_{\epsilon}A^{(3)}&=&-{\rm d}\epsilon^{(3)}
+[A^{(0)},\epsilon^{(3)}]+[A^{(2)},\epsilon^{(1)}]\, ,\\\nonumber
 \delta_{\epsilon}A^{(2)}&=&
[A^{(1)},\epsilon^{(1)}]+[A^{(3)},\epsilon^{(3)}]\,  ,\eea where
we have assumed that $\epsilon=\epsilon^{(1)}+\epsilon^{(3)}$.
Using these formulae we find for the variation of the Lagrangian
density  \bea \nonumber
\delta_{\epsilon}\L&=&\delta\gamma^{\a\b}{\rm
str}\big(A^{(2)}_{\a}A^{(2)}_{\b}\big)-2\gamma^{\a\b}{\rm
str}\big([A^{(1)}_{\a},A^{(2)}_{\b}]\epsilon^{(1)}
+[A^{(3)}_{\a},A^{(2)}_{\b}]\epsilon^{(3)}\big)\\
\nonumber &+&\kappa \epsilon^{\a\beta}{\rm
str}\Big(\pa_{\a}A^{(3)}_{\b}\epsilon^{(1)}
-\pa_{\a}A^{(1)}_{\beta}\epsilon^{(3)}+[A^{(0)}_{\a},\epsilon^{(1)}]A^{(3)}_{\b}
+[A^{(2)}_{\a},\epsilon^{(3)}]A^{(3)}_{\b}
\\
&&~~~~~~~~~~~~~~ +A^{(1)}_{\a}[A^{(0)}_{\b},\epsilon^{(3)}]
+A^{(1)}_{\a}[A^{(2)}_{\b},\epsilon^{(1)}] \Big)\, . \eea Here we
used integration by parts to remove the derivatives of $\epsilon$.
The variation of the world-sheet metric is left unspecified. Now
we note that the zero curvature condition eq.(\ref{zeroc}) implies
\bea \nonumber \epsilon^{\a\b} \pa_{\a}
A_{\b}^{(1)}&=&\epsilon^{\a\b}[A_{\a}^{(0)},A_{\b}^{(1)}]
+\epsilon^{\a\b}[A_{\a}^{(2)},A_{\b}^{(3)}] \, ,\\
\nonumber\epsilon^{\a\b} \pa_{\a}
A_{\b}^{(3)}&=&\epsilon^{\a\b}[A_{\a}^{(0)},A_{\b}^{(3)}]
+\epsilon^{\a\b}[A_{\a}^{(1)},A_{\b}^{(2)}]  \, . \eea Taking this
into account, we obtain $$
\delta_{\epsilon}\L=\delta\gamma^{\a\b}{\rm
str}\big(A^{(2)}_{\a}A^{(2)}_{\b}\big)-4\, {\rm str}\big( {\rm
P}_{+}^{\a\b}[A^{(1)}_{\b},A^{(2)}_{\a}]\epsilon^{(1)} +{\rm
P}_{-}^{\a\b}[A^{(3)}_{\b},A^{(2)}_{\a}]\epsilon^{(3)} \big) \, .
$$
According to this formula, the variation of the Lagrangian
trivially vanishes for field configurations which solve equations
of motion (\ref{Eqf}) and the Virasoro constraints (\ref{Vir}). In
particular, the variation of the first term is zero due to the
identity $\gamma_{\a\b}\delta\gamma^{\a\b}=0$ which follows from
the condition $\det \gamma=-1$. Of course, under $\kappa$-symmetry
transformations the action should remain invariant {\it without}
using equations of motion.
\smallskip

In what follows we assume that $\k=\pm 1$. For any vector $V^{\a}$ we introduce the projections $V^\a_\pm$:
$$
V^\a_\pm = {\rm P}_{\pm}^{\a\b}V_\b\,
$$
so that the variation of the Lagrangian acquires the form
$$
\delta_{\epsilon}\L=\delta\gamma^{\a\b}{\rm
str}\big(A^{(2)}_{\a}A^{(2)}_{\b}\big)-4\, {\rm str}\big(
[A^{(1),\a}_{+},A^{(2)}_{\a,-}]\epsilon^{(1)}
+[A^{(3),\a}_{-},A^{(2)}_{\a,+}]\epsilon^{(3)} \big) \, .$$ We
further note that from the condition ${\rm
P}_{\pm}^{\a\b}A_{\b,\mp}=0$ the components $A_{\tau,\pm}$ and
$A_{\sigma,\pm}$ are proportional \bea \label{prop}
A_{\tau,\pm}=-\frac{\gamma^{\tau\sigma}\mp\kappa
}{\gamma^{\tau\tau}}A_{\sigma,\pm}\, . \eea To simplify our
further treatment, we put for the moment $\epsilon^{(3)}=0$.

\smallskip

The crucial point of our construction is the following ansatz for
the $\kappa$-symmetry parameter $\epsilon^{(1)}$ \bea
\hspace{-0.7cm}\epsilon^{(1)}=A_{\a,-}^{(2)}A_{\b,-}^{(2)}\kappa_{++}^{\a\b}
+\kappa_{++}^{\a\b}A_{\a,-}^{(2)}A_{\b,-}^{(2)}+A_{\a,-}^{(2)}\kappa_{++}^{\a\b}A_{\b,-}^{(2)}-\frac{1}{8}\,
{\rm str}(\Sigma A_{\a,-}^{(2)}A_{\b,-}^{(2)})\kappa_{++}^{\a\b}\,
, \label{kappa} \eea where $\kappa_{++}^{\a\b}$ is the
$\kappa$-symmetry parameter which is assumed to be independent on
the dynamical fields of the model.
  Obviously, $\kappa^{\a\b}_{++}$ must be an
element of $\alg{ osp}(2,2|6)$. Since on the product of any three
supermatrices $A,B$ and $C$ the automorphism $\Omega$ acts as
$\Omega(ABC)=\Omega(C)\Omega(B)\Omega(A)$, we see that
$\epsilon^{(1)}\in \A^{(1)}$ provided $\kappa^{\a\b}_{++}$ is also
an element of degree one: $\kappa^{\a\b}_{++}\in \A^{(1)}$.

\smallskip

Consider now the commutator
 \bea\nonumber [A^{(2)}_{\a,-},\epsilon^{(1)}]
 &=& A^{(2)}_{\a,-}A_{\b,-}^{(2)}A_{\delta,-}^{(2)}\kappa_{++}^{\b\delta}
+A^{(2)}_{\a,-}\kappa_{++}^{\b\delta}A_{\b,-}^{(2)}A_{\delta,-}^{(2)}+A^{(2)}_{\a,-}A_{\b,-}^{(2)}
\kappa_{++}^{\b\delta}A_{\delta,-}^{(2)}\\
\nonumber
 &-&
A_{\b,-}^{(2)}A_{\delta,-}^{(2)}\kappa_{++}^{\b\delta}A^{(2)}_{\a,-}
-\kappa_{++}^{\b\delta}A_{\b,-}^{(2)}A_{\delta,-}^{(2)}A^{(2)}_{\a,-}
-A_{\b,-}^{(2)}
\kappa_{++}^{\b\delta}A_{\delta,-}^{(2)}A^{(2)}_{\a,-}\\
 &-&\frac{1}{8}\,
{\rm str}(\Sigma A_{\b,-}^{(2)}A_{\delta,-}^{(2)})
A^{(2)}_{\a,-}\kappa_{++}^{\b\delta}+ \frac{1}{8}\, {\rm
str}(\Sigma A _{\b,-}^{(2)}A_{\delta,-}^{(2)})
\kappa_{++}^{\b\delta}A^{(2)}_{\a,-}\, . \label{com}\eea Here we
have to deal with tensorial structures
$$
A^{(2)}_{\a,-}\ldots A_{\b,-}^{(2)}\ldots A_{\delta,-}^{(2)}\,,
$$
where dots indicate insertions of other supermatrices, e.g.,
$\kappa_{++}$. Since $A^{(2)}_{\a,-}$ is an (anti-)self-dual form,
the tensors above are totally symmetric in indices $\a,\b,\delta$
and have, in fact, a single non-trivial entry, all the other
entries being proportional to it. Thus, most of the terms in
eq.(\ref{com}) are cancelled out and we are left with \bea
[A^{(2)}_{\a,-},\epsilon^{(1)}]
 &=&[A^{(2)}_{\a,-}A_{\b,-}^{(2)}A_{\delta,-}^{(2)}-\sfrac{1}{8}\,
{\rm str}(\Sigma A_{\b,-}^{(2)}A_{\delta,-}^{(2)})
A^{(2)}_{\a,-},\kappa_{++}^{\b\delta}]\, . \eea Now we invoke the
identity (\ref{identity}) satisfied by any element $A\in\A^{(2)}$,
according to which
$$
A^{(2)}_{\a,-}A_{\b,-}^{(2)}A_{\delta,-}^{(2)}-\frac{1}{8}\, {\rm
str}(\Sigma A_{\b,-}^{(2)}A_{\delta,-}^{(2)})
A^{(2)}_{\a,-}=\frac{1}{8}\, {\rm
str}(A_{\b,-}^{(2)}A_{\delta,-}^{(2)})\Sigma
 A^{(2)}_{\a,-}\, .
$$
Thus, we have found that
\bea [A^{(2)}_{\a,-},\epsilon^{(1)}]
 &=& \frac{1}{8}\, {\rm str}(A_{\b,-}^{(2)}A_{\delta,-}^{(2)})[\Sigma
 A^{(2)}_{\a,-},\kappa_{++}^{\b\delta}]\, .
\eea Now we see that the $\kappa$-symmetry variation of the action
\bea \delta_{\epsilon}\L=\delta\gamma^{\a\b}{\rm
str}\big(A^{(2)}_{\a}A^{(2)}_{\b}\big)-4\, {\rm str}\Big(
[A^{(1),\a}_{+},A^{(2)}_{\a,-}]\epsilon^{(1)} \Big) \,
 \eea
implies then the following transformation law for the
two-dimensional metric \bea \delta \gamma^{\a\beta}=\frac{1}{2}\,
{\rm str}\Big(\Sigma A^{(2)}_{\delta,-}
[\kappa_{++}^{\a\b},A^{(1),\delta}_+]\Big)\,. \eea Notice that the
condition $\gamma_{\a\beta}\delta \gamma^{\a\beta}$ is
automatically obeyed because
$$
\gamma_{\a\beta}\delta \gamma^{\a\beta}
=\gamma^{\a\b}P^+_{\a\delta}P^+_{\beta\eta}\kappa^{\delta\eta}=0\,
.
$$
Using the fact that the matrix $\Sigma$ anti-commutes with any
fermionic matrix, we can rewrite the $\kappa$-variation of the
metric as \bea \delta \gamma^{\a\beta}=\frac{1}{2}\, {\rm
str}\Big(\Sigma \kappa_{++}^{\a\b} \{A^{(1),\delta}_+,
A^{(2)}_{\delta,-}\} \Big)\, . \eea We see that in a certain sense
the variation occurs in the direction orthogonal to the fermionic
equations of motion which are $[A^{(1),\delta}_+,
A^{(2)}_{\delta,-}]=0$.

\smallskip
It is obvious that the treatment above can be repeated for the
variation involving $\epsilon^{(3)}$, so that a complete variation
of the metric under $\kappa$-symmetry will be of the form \bea
\delta \gamma^{\a\beta}=\frac{1}{2}\, {\rm str}\Big(\Sigma
A^{(2)}_{\delta,-}
[\kappa_{++}^{\a\b},A^{(1),\delta}_+]\Big)+\frac{1}{2}\, {\rm
str}\Big(\Sigma A^{(2)}_{\delta,+}
[\varkappa_{--}^{\a\b},A^{(3),\delta}_-]\Big)\, , \eea where
$\varkappa_{--}^{\a\b}\subset \A^{(3)}$ is another independent
$\kappa$-symmetry parameter.

\smallskip

We would like to point out that in our derivation of
$\kappa$-symmetry we used the fact that ${\rm P}_{\pm}^{\a\b}$ are
orthogonal projectors and, therefore, realization of the
$\kappa$-symmetry requires the parameter $\kappa$ in the
Lagrangian to be equal to $\pm 1$.

\subsection{Rank of $\kappa$-symmetry transformations on-shell}

The next important question is to understand how many fermionic
degrees of freedom can be gauged away on-shell by means of
$\kappa$-symmetry. To this end one can make use of the light-cone
gauge. Generically, the light-cone coordinates $X^{\pm}$ are
introduced by making linear combinations of one field
corresponding to the time direction from ${\rm AdS}_4$ and one
field from $\CP^3$. Without loss of generality we can assume that
the transversal fluctuation are all suppressed and the
corresponding element $A^{(2)}$ has the form \bea\label{Alc}
A^{(2)}= \left(\begin{array}{cc} i x\Gamma^0 & 0
\\
0 & y T_6
\end{array} \right)\, .
\eea Indeed, the matrix $\Gamma^0$ corresponds to the time
direction in ${\rm AdS}_4$ and any element from the tangent space
to $\CP^3$ can be brought to $T_6$ by means of an $\alg{so}(6)$
transformation. The Virasoro constraint ${\rm
str}(A^{(2)}_{\a,-}A^{(2)}_{\b, -})=0$ then demands that
$x^2=y^2$, {\it i.e.} $x=\pm y$. Picking up, e.g., the first
solution $x=y$, we then compute the element $\epsilon^{(1)}$
assuming a generic parameter $\kappa$, which depends on 12
independent fermionic variables.\footnote{The matrix $\kappa_{++}$
depends on 12 fermionic variables only, because it is an element
of $\A^{(1)}$. } First, we see that ${\rm str}(\Sigma
A^{(2)}A^{(2)})=-8x^2$. Second, plugging eq.(\ref{Alc}) into
eq.(\ref{kappa}), we obtain \bea
\epsilon^{(1)}=x^2\left(\begin{array}{cc} 0 ~&~ \varepsilon
\\
-\varepsilon^tC_4 ~&~ 0
\end{array} \right)\, , \eea
where $\varepsilon$ is the following matrix \bea
\nonumber\varepsilon={\small \left(\begin{array}{rrrrrr}
 0 ~&~ 0 ~&~ i(i k_{13}-k_{16}) ~&~ i(ik_{14}-k_{15}) ~&~ ik_{14}-k_{15} ~&~
i k_{13}-k_{16}
\\
 0 ~&~ 0 ~&~ i(ik_{23}-k_{26}) ~&~ i(ik_{24}-k_{26}) ~&~ ik_{24}-k_{25} ~&~ ik_{23}-k_{26}\\
 0 ~&~ 0 ~&~-i(-ik_{33}-k_{36}) ~&~ -i(-ik_{34}-k_{35}) ~&~ -ik_{34}-k_{35} ~&~
-ik_{33}-k_{36}  \\
 0 ~&~ 0 ~&~ -i(-ik_{43}-k_{46}) ~&~-i( -ik_{44}-k_{45}) ~&~ -ik_{44}-k_{45} ~&~
-ik_{43}-k_{46}
\end{array} \right)\,
} \eea and $\kappa_{ij}\equiv \kappa_{++,ij}$ are the entries of
the matrix $\kappa_{++}$. As we see, the matrix $\varepsilon$
depends on 8 independent complex fermionic parameters (e.g. the
last two columns). The reality condition (\ref{rel3}) for
$\varepsilon$ reduces this number by half. Finally,
$\epsilon^{(1)}$ must belong to the component $\A^{(1)}$ which
further reduce the number of fermions by half. As the result,
$\epsilon^{(1)}$ depends on four real fermionic parameters. A
similar analysis shows that $\epsilon^{(3)}$ will also depend on
four real fermions. Thus, in total $\epsilon^{(1)}$ and
$\epsilon^{(3)}$ depend on 8 real fermions and these are those
degrees of freedom which can be gauged away by $\kappa$-symmetry.
The gauge-fixed coset model will therefore  involve 16 physical
fermions only.

\smallskip

It should be noted that the considerations above are applicable to
a generic case, where string motion occurs in both ${\rm AdS}_4$
and $\CP^3$ spaces. There is however a singular situation, when
string moves in the AdS space only (e.g. the  string spinning in
$\ads_3$ \cite{Gubser:2002tv}). One can show that for this case
the transformation (\ref{kappa}) vanishes, although the fermionic
equations remain degenerate and only 12 of them (out of 24) are
independent. This suggests that realization of $\kappa$-symmetry
changes in this singular situation and $\kappa$-symmetry  becomes
capable of gauging away 12 from 24 fermions. A singular nature of
the corresponding bosonic background shows up in the fact that as
soon as fluctuations along $\CP^3$ directions are switched on, the
rank of $\kappa$-symmetry gets reduced to 8. As the result,
singular backgrounds cannot be quantized semi-classically within
the coset sigma-model. This picture is rather different from that
for conventional type IIA or IIB superstrings. There
$\kappa$-symmetry can always remove half of fermionic degrees of
freedom. Which half however, does depend on a chosen bosonic
background. As was already explained above, we would like to treat
our coset model as the one which originates from the type IIA
superstring on ${\rm AdS}_4\times \CP^3$ upon a partial
$\kappa$-symmetry fixing. The trouble with singular (AdS)
backgrounds we observe here could be, therefore, due to their
incompatibility with the $\kappa$-symmetry gauge choice which
reduces type IIA superstring on ${\rm AdS}_4\times \CP^3$ to our
coset model. It would be important to further clarify this issue.

\section{Integrability}
Since there is no difference in construction of the Lagrangian in
comparison to the case of $\AdS$, the Lax connection found in
\cite{Bena:2003wd} for superstrings on $\AdS$ is applicable to our
model as well and, therefore, we can straightforwardly conclude
its kinematical integrability. The main result of this section
consists in showing that under $\kappa$-symmetry variations found
in the previous section the Lax connection undergoes a gauge
transformation on-shell. This provides another non-trivial check
that $\kappa$-symmetry transforms solution of the equations of
motion into solutions.

\subsection{Lax connection}
The Lax representation of the superstring equations of motion on
$\AdS$ has been found in \cite{Bena:2003wd}. The corresponding
two-dimensional Lax connection ${L}_{\a}$ has the following
structure
 \bea
 \label{wL}
{L}_{\a}=\ell_0 A_{\a}^{(0)}+\ell_1 A_{\a}^{(2)}
+\ell_2\gamma_{\a\b}\epsilon^{\beta\rho}A_{\rho}^{(2)} +\ell_3
A_{\a}^{(1)}+\ell_4 A_{\a}^{(3)}\, , \eea where $\ell_i$ are
some constants.

\smallskip

The connection ${L}$ is required to have zero curvature as a
consequence of the dynamical equations and the flatness of
$A_{\a}$. This requirement allows one to determine the constants
$\ell_i$. For the reader's convenience below we summarize the
result for $\ell_i$.

\smallskip

All the parameters $\ell_i$ are determined in terms of $\ell_1$:
\bea \ell_3^2=\ell_1\pm \sqrt{\ell_1^2-1}\,
,~~~~~\ell_4^2=\ell_1\mp \sqrt{\ell_1^2-1}\, , ~~~~~\ell_2=\pm
\sqrt{\ell_1^2-1}\,  , ~~~\ell_0=1\, .\eea The signs in these
formulae correlate with the corresponding sign of $\kappa$ which
is also required to satisfy the condition $\kappa^2=1$.
 It is convenient to
describe all the coefficients in terms of uniformizing {spectral}
parameter $z$. We parametrize
$$\ell_1=\frac{1+z^2}{1-z^2}\, . $$
For the remaining coefficients $\ell_i$ the complete set of
solutions reads as follows\bea \label{fp1}
\ell_2&=&s_1\frac{2z}{1-z^2}\, , ~~~~~\ell_3=s_2\frac{1+s_2s_3
z}{\sqrt{1-z^2}}\, ,~~~~~~~\ell_4=s_2\frac{1-s_2s_3
z}{\sqrt{1-z^2}} \, , \eea Here $s_2^2=s_3^2=1$ and $s_1s_2s_3=-{\rm sign}\,\kappa$.
 Thus, for every
choice of $\kappa$ we have four different solutions for $\ell_i$
specified by the choice of $s_2=\pm 1$ and $s_3=\pm 1$. The
spectral parameter $z$ takes values in the complex plane and, for
this reason, the Lax connection takes values in the complexified
algebra $\alg{osp}(4|6)$.

\smallskip

Finally, we point out how the grading map $\Omega$ acts on the Lax
connection $L_{\a}$. Since $\Omega$ is the automorphism of $\osp$
the curvature of $\Omega(L_{\a})$ also vanishes. It can be easily
checked that $\Omega(L_{\a})$ is related to $L_{\a}$ by a certain
diffeomorphism of the spectral parameter, namely,
$$
\Omega(L_{\a}(z))=\Upsilon L_{\a}(z)\Upsilon^{-1}=L_{\a}(1/z)\, .
$$

\smallskip

In summary, equations of motion admit the same zero-curvature
representation as for superstring on  $\AdS$ which ensures the
kinematical integrability of our coset model. Inclusion of the
Wess-Zumino term is allowed by integrability only for $\kappa=\pm
1$, {\it i.e.} only for those values of $\kappa$ for which the
model has the local fermionic symmetry.

\subsection{$\kappa$-variations of the Lax connection}
In this section we analyse the relationship between the Lax
connection and $\kappa$-symmetry. By using the formulae
(\ref{dcst}) which describe how the $\mathbb{Z}_4$-components of
$A=-g^{-1}{\rm d }g$ transform under the $\kappa$-symmetry, it is
straightforward to find the $\kappa$-symmetry variation of the Lax
connection \bea \delta L_{\a}=[L_{\a},\Lambda]-\pa_{\a}\Lambda
+\ell_2\ell_3\epsilon_{\a\b}[A^{(2),\beta}_-,\epsilon^{(1)}] +
\ell_2\epsilon_{\a\beta}\Big(2[A^{(1),\beta}_{+},\epsilon^{(1)}]+\delta\gamma^{\beta\delta}A_{\delta}^{(2)}\Big)\,
,\eea where $\Lambda=\ell_3 \epsilon^{(1)}$. The last two terms
proportional to $\ell_2\ell_3$ and $\ell_2$ would destroy the zero
curvature condition for the $\kappa$-transformed connection unless
they separately vanish. Concerning the first term, as we have
shown in the previous section, \bea
[A^{(2)}_{\a,-},\epsilon^{(1)}]
 &=& \frac{1}{8}{\rm str}(A_{\b,-}^{(2)}A_{\delta,-}^{(2)})[\Sigma
 A^{(2)}_{\a,-},\kappa_{++}^{\b\delta}]\, ,\eea so that this term
vanishes due to the Virasoro constraints:
$$
{\rm str}(A_{\a,-}^{(2)}A_{\b,-}^{(2)})=0\, .
$$
As to the second term, by using equations of motion for fermions
the relevant commutator  can be written as follows
\bea \nonumber
[A^{(1),\beta}_{+},\epsilon^{(1)}]&=&
A_{\a,-}^{(2)}A_{\b,-}^{(2)}[A^{(1),\beta}_{+},\kappa_{++}^{\a\b}]
+[A^{(1),\beta}_{+},\kappa_{++}^{\a\b}]A_{\a,-}^{(2)}A_{\b,-}^{(2)}\\
&+&A_{\a,-}^{(2)}[A^{(1),\beta}_{+},\kappa_{++}^{\a\b}]A_{\b,-}^{(2)}-\frac{1}{8}\,
{\rm str}(\Sigma
A_{\a,-}^{(2)}A_{\b,-}^{(2)})[A^{(1),\beta}_{+},\kappa_{++}^{\a\b}]\,
. \label{com1} \eea We assume a parametrization \bea A^{(2)}=
\left(
\begin{array}{cc}
y_{\mu} \Gamma^{\mu} & 0 \\ 0 & \bar{y}_i T_{i}
\end{array}
\right) \, , ~~~~~~~~~~[A^{(1),\beta}_{+},\kappa_{++}^{\a\b}]=
\left(
\begin{array}{cc}
u_{\mu} \Gamma^{\mu} & 0 \\ 0 & \bar{u}_i T_{i}
\end{array}
\right) \, \eea and, therefore, $\frac{1}{8}\, {\rm str}(\Sigma
A_{\a,-}^{(2)}A_{\b,-}^{(2)})=\frac{1}{2}(y^2-\bar{y}^2)$, while
the Virasoro constraint is \bea\label{vc}
{\rm
str}(A_{\a,-}^{(2)}A_{\b,-}^{(2)})=4(y^2+\bar{y}^2)=0\,
~~~\Longrightarrow~~~y^2=-\bar{y}^2\, . \eea
 With
this parametrization at hand,  the r.h.s of eq.(\ref{com1}) boils
down to the following matrix expression {\small \bea
[A^{(1),\beta}_{+},\epsilon^{(1)}]&=&\left(
\begin{array}{cc}
y_{\mu}y_{\nu}u_{\rho}
\Big(\Gamma^{\mu}\Gamma^{\nu}\Gamma^{\rho}+\Gamma^{\mu}\Gamma^{\rho}\Gamma^{\nu}+\Gamma^{\rho}\Gamma^{\mu}
\Gamma^{\nu}\Big) & 0 \\
0 & \bar{y}_i\bar{y}_j\bar{u}_k
\Big(T_{i}T_jT_k+T_{i}T_kT_j+T_kT_iT_j\Big)
\end{array}
\right)  \nonumber \\
&-&\frac{1}{2}(y^2-\bar{y}^2)\left(
\begin{array}{cc} u_{\mu}\Gamma^{\mu} & 0 \\ 0 & \bar{u}_iT_i\end{array}\right)\, .\nonumber \eea
} The Clifford algebra for the gamma-matrices together with the
permutation properties for $T_i$'s allows one to rewrite the above
formula as \bea  [A^{(1),\beta}_{+},\epsilon^{(1)}]&=&
{\small\left(
\begin{array}{cc}
y^2\, u_{\mu}\Gamma^{\mu} +2(yu)\,  y_{\mu}\Gamma^{\mu}  & 0 \\
0 & -\bar{y}^2\,  \bar{u}_i T_i -2 (\bar{y}\bar{u})\, \bar{y}_i
T_i
\end{array}
\right)-\frac{1}{2}(y^2-\bar{y}^2)\left(
\begin{array}{cc} u_{\mu}\Gamma^{\mu} & 0 \\ 0 & \bar{u}_iT_i\end{array}\right) \, , } \nonumber \eea
where we defined $(yu)\equiv y_{\mu}u_{\nu}\eta^{\mu\nu}$ and
$(\bar{y}\bar{u})\equiv \bar{y}_i\bar{u}_i$. Taking into account
the Virasoro constraints, the last expression simplifies to \bea
[A^{(1),\beta}_{+},\epsilon^{(1)}]&=&  \left(
\begin{array}{cc}
2(yu)\,  y_{\mu}\Gamma^{\mu}  & 0 \\
0 &  - 2(\bar{y}\bar{u})\, \bar{y}_i T_i
\end{array}
\right)\, .\eea We also notice that due to the fermion equations
of motion the products $(yu)$ and $(\bar{y}\bar{u})$ are not
independent. Indeed, on-shell we have \bea 0={\rm
str}([A^{(1),\a}_{+},\epsilon^{(1)}]A^{(2)}_{\a,-}) =
8(y^2(yu)-\bar{y}^2\big(\bar{y}\bar{u})\big)\, .\eea The Virasoro
constraints (\ref{vc}) then imply that \bea
(yu)=-(\bar{y}\bar{u})\, . \eea On the other hand, \bea
\delta\gamma^{\beta\delta}A_{\delta}^{(2)}=-2((yu)-(\bar{y}\bar{u}))
\left(
\begin{array}{cc}
y_{\mu} \Gamma^{\mu} & 0 \\ 0 & \bar{y}_i T_{i}
\end{array}
\right)=-4(yu)\left(
\begin{array}{cc}
y_{\mu} \Gamma^{\mu} & 0 \\ 0 & \bar{y}_i T_{i}
\end{array}
\right) \, . \eea Thus, \bea
2[A^{(1),\beta}_{+},\epsilon^{(1)}]+\delta\gamma^{\beta\delta}A_{\delta}^{(2)}=0\,
, \eea {\it i.e.} a $\kappa$-symmetry variation of the Lax
connection is a gauge transformation on-shell.

\section{Plane-wave limit}

In this section we discuss the perturbative expansion of the
string sigma model action up to the quadratic order in the bosonic
and fermionic fields  around a point-like string solution
describing a massless particle moving in $\CP^3$ along a null
geodesic given by  the equations $w_1=w_2=0\,,\ w_3 = e^{i\phi}$.
The reader should consult Appendix \ref{apar} for notations and
parametrizations of $\CP^3$ used in the paper. The expansion
corresponds to taking a Penrose  or plane-wave limit of the
background $\ads_4\times\CP^3$ geometry which was recently
discussed in \cite{Nishioka:2008gz,Gaiotto:2008cg}. The resulting
type IIA string theory  pp-wave  background has 24
supersymmetries, and the corresponding  light-cone gauge
Green-Schwarz action describes 8 massive bosons and 8 massive
fermions, and  was constructed in \cite{Sugiyama:2002tf,IIAlc}. We
use the sigma model action (\ref{stac}) to compute the quadratic
action, then we impose a certain $\kappa$-symmetry gauge condition
and show that the light-cone gauge action coincides with the one
in \cite{Sugiyama:2002tf,IIAlc}. We consider this computation as a
first nontrivial check of our coset sigma model action for
superstrings on $\ads_4\times\CP^3$.

\smallskip

To find a reasonably good expansion around the geodesics, it is
convenient to use the homogeneous coordinates $z_i$ of $\CP^3$.
Then one can see that the parametrization of $z_i$  which leads to
a simple bosonic quadratic action describing massive excitations
can be chosen as follows \bea \nonumber z_4 = e^{-i\phi/2}\,,\quad
z_3 = (1 - x_4)e^{i\phi/2}\,,\quad z_1 = {1\ov\sqrt 2} y_1\,,\quad
z_2 = {1\ov\sqrt 2} y_2\,, \eea where the angle $\phi$
parametrizes the geodesics, and the complex coordinates $y_1,y_2$
and the real coordinate $x_4$ denote the five physical
fluctuations in $\CP^3$. In terms of the inhomogeneous coordinates
$w_i$ the parametrization has the form \bea\nonumber w_3 = (1 -
x_4)e^{i\phi}\,,\quad w_1 = {1\ov\sqrt 2} y_1e^{i\phi/2}\,,\quad
w_2 = {1\ov\sqrt 2} y_2e^{i\phi/2}\,, \eea and all the coordinates
$w_i$ depend on $\phi$.

\smallskip

Then,
expanding the $\CP^3$ metric (\ref{metricz})  in powers of $x_4, y_1, y_2$, one
finds \bea\nonumber 4\, ds^2_{_{\CP^3}} =  d\phi^2( 1 - x_4^2 -  {1\ov
4}{\bar y}_r y_r)+ dx_4^2 + d{\bar y}_r dy_r  + \cdots\,,\quad
r=1,2\,. \eea
This formula should be combined with the standard expansion
of the $\ads_4$ metric, see e.g. \cite{Arutyunov:2004yx} for a convenient parametrization of $\ads_d$
\bea\nonumber
 ds^2_{_{\ads_4}} =  - dt^2( 1 + x_i^2 ) + dx_i^2  + \cdots\,,\quad i=1,2,3\,,
\eea
where $t$ is the global time coordinate, and  $x_i$ are three physical fluctuations in $\ads_4$.

\smallskip

Thus,  the $\ads_4\times\CP^3$
background metric admits the following expansion
\bea\la{expmet}
 ds^2_{_{\ads_4\times\CP^3}}=  - dt^2( 1 + x_i^2 ) + dx_i^2  +d\phi^2( 1 - x_4^2 -  {1\ov 4}{\bar y}_r y_r)+ dx_4^2 + d{\bar y}_r dy_r  + \cdots\,.~~~~
\eea It is clear now that plugging in the point-like string
solution with $t=\tau\,,\ \phi =\tau$ in the corresponding string
Lagrangian (\ref{sLag}) one gets four massive fields of mass 1/2 and
four fields of mass 1. Note that the field $x_4$ from $\CP^3$
joins the three fields from $\ads_4$. It is unclear at the moment
if it is a consequence of the supersymmetry or an artifact of the
plane-wave expansion and sigma-model loop corrections would result
in a mass splitting.

\bigskip

To find the quadratic fermion action in the background, we need to
know the coset representative corresponding to the point-like
string solution. Since for the solution $w_1=w_2=0\,,\ w_3 =
e^{i\phi}$, it is given by \bea\la{gads0}
g_{B} &=&\left(\begin{array}{rr} g_{_{\ads}}& 0  \\
0~~ &~ g_{_{\CP}}
\end{array}\right)\,, \quad g_{_{\ads}} = e^{i t \Gamma^0/2}  \,, \quad \\\la{gcp0}
g_{_{\CP}} &=&  I + { e^{i\phi}{\cal T}_3 + e^{-i\phi}{\overline
{\cal T}}_3\ov\sqrt{2}}  + (1 - {1\ov \sqrt{2}}) ({\cal T}_3
{\overline {\cal T}}_3 + {\overline{\cal T}}_3{\cal T}_3)\,, \eea
where we use (\ref{gcos}), and take into account that the time direction corresponds to $\Gamma^0$.

\smallskip

Then we build up the group element in the form \bea\la{cosg}  g =
g(\chi)\, g_B\,  \eea and compute the quadratic part of the
fermion Lagrangian (\ref{sLag}). In the last formula
$g(\chi)=\exp\chi$, where $\chi$ is a generic odd element of
$\osp$.

\smallskip

One can check that  the coset representative $g_{_{\CP}}$ (\ref{gcp0}) does not correspond to any one-parameter subgroup of SO(6) because the tangent element to
 $g_{_{\CP}}$ has an explicit dependence on $\phi$. For this reason
it seems easier to parametrize the coset manifolds $\ads_4$ and $\CP^3$ by the following group elements
 \bea \la{Gg}  G= {\rm
diag}(G_{_{\ads}} ,\ G_{_{\CP}} )\,,\quad  G_{_{\ads}}  =
g_{_{\ads}} \, K_4\, g_{_{\ads}}^t\,,\quad G_{_{\CP}}  =
g_{_{\CP}} \, K_6\, g_{_{\CP}}^t\, . \eea Indeed, as was shown in
\cite{Alday:2005ww}, the sigma model string Lagrangian of the form
(\ref{sLag}) can be rewritten in terms of these elements as
follows
 \bea\nonumber \L&=&{1\ov 4}\, {\rm
str}\left[\gamma^{\a\b} (\B_{\a}+G\B_{\a}^tG^{-1}+\pa_{\a}G
G^{-1})(\B_{\b}+G\B_{\b}^tG^{-1}+\pa_{\b}G
G^{-1})\right.\\\la{Lor} &&~~~~~~~~~~~~~+\left.
2i\kappa\epsilon^{\a\b}{\rm F}_{\a} G{\rm F}_{\b}^{st}
G^{-1}\right]\, , \eea where ${\rm F}$ and ${\rm B}$ are odd and
even superalgebra elements made of fermions only \bea \label{gdg}
g^{-1}(\chi)dg(\chi) ={\rm F} + {\rm B}\,,\quad {\rm F} = d\chi +
\cdots\,,\quad {\rm B}= {1\ov 2}d\chi\chi -{1\ov 2}\chi d\chi
+\cdots\, . \eea The coset group elements are skew-symmetric
matrices $G^t = -G$, and, therefore, $\ads_4$ can be identified
with the intersection of $4\times 4$ skew-symmetric matrices with
USP(2,2) ones, and $\CP^3$ with the intersection of $6\times 6$
skew-symmetric and orthogonal matrices.  The parametrizations
(\ref{cosg}) of the supergroup elements  and (\ref{Gg}) of the
coset manifolds are distinguished because the bosonic subgroup of
OSP$(2,2|6)$ acts on the coset representatives $g(\chi)$ and $G$
by the usual matrix conjugation \cite{AAF1}.

\smallskip

The string Lagrangian (\ref{Lor}) can be further simplified by
taking into account that \bea\nonumber {\rm str}\left(
G\B_{\a}^tG^{-1}\pa_{\b}G G^{-1}\right)= {\rm str}\left(
\B_{\a}\pa_{\b}G G^{-1}\right)\, , \eea and, therefore, we can
bring  (\ref{Lor}) to the following simple form
 \bea\la{Lor2}  \L={1\ov 4}{\rm
str}\left[\gamma^{\a\b} (\pa_{\a}GG^{-1}\pa_{\b}GG^{-1} + 4
\B_{\a}\pa_{\b}G G^{-1}+\B_{\a}\B_{\b})+ 2i\kappa\epsilon^{\a\b}{\rm
F}_{\a} G{\rm F}_{\b}^{st} G^{-1}\right].~~~~~ \eea
The first term in the expression gives the bosonic part of the string Lagrangian determined by the background metric. According to formula (\ref{expmet}), the corresponding action expanded around the particle trajectory $t=\tau\,,\ \phi=\tau$ takes the following form
\bea
\la{sbos2}
 S_B^{(2)}=-\frac{R^2}{4\pi \a'}\int_0^{{\cal J}} {\rm d}\sigma{\rm d}\tau\,
 \Big( \pa^\a x_k\pa_\a x_k  -   x_k^2 + \pa^\a {\bar y}_r\pa_\a y_r   -  {1\ov 4}{\bar y}_r y_r \Big)\,,
\eea
where $k=1,2,3,4$, the integration limit $ {\cal J}$ is determined by the charge (or target space-time energy) carried by the particle: $E=J = \frac{R^2}{2\pi \a'} {\cal J}$, and we dropped the unessential fluctuations in the time and $\phi$ directions.

\smallskip

The quadratic fermion action in the particle background is given by the sum of the second  and fourth terms in (\ref{Lor2})
 \bea\la{Lorf2}  \L_F^{(2)}={\rm
str}\left[\gamma^{\a\b}
\B_{\a}\pa_{\b}G G^{-1}+ {i\ov 2}\kappa\epsilon^{\a\b}{\rm
F}_{\a} G{\rm F}_{\b}^{st} G^{-1}\right]\,,~~~~~ \eea
where
 \bea \label{gdg2}
 {\rm F} = d\chi\,,\quad {\rm B}=
{1\ov 2}d\chi\chi -{1\ov 2}\chi d\chi \, , \eea
and $G$ is given by (\ref{Gg}), (\ref{gads0}) and (\ref{gcp0}) with $t=\phi=\tau$.

\smallskip

 Then, one can show that $G_{_{\CP}}$ with  $g_{_{\CP}}$ given by eq.(\ref{gcp0})
 admits the following simple representation
\bea\nonumber G_{_{\CP}}  = g_{_{\CP}} \, K_6\, g_{_{\CP}}^t = (I
+ e^{i\phi}{\cal T}_3 + e^{-i\phi}{\overline {\cal T}}_3  + {\cal
T}_3 {\overline {\cal T}}_3 + {\overline{\cal T}}_3{\cal T}_3)K_6
= h(\phi)\, K_6\, h(\phi)^t\,,~~~~~~ \eea where \bea\nonumber
h(\phi) = e^{-\phi T_{56}} e^{{\pi\ov 2}T_{35}}\,,\quad T_{ij} =
E_{ij}-E_{ji}\,. \eea Therefore, we can redefine the fermions and
bosons as \bea \nonumber G\to HGH^{t}\,, \quad\chi\to H\, \chi \,
H^{-1}\,,\quad H = {\rm diag } ( g_{_{\ads}}(t) ,\ h(\phi) )\,,
\eea and remove the explicit dependence of $t$ and $\phi$ from the
Lagrangian leaving only the dependence of their derivatives. One
can easily see that the redefinition amounts to the following
replacement in the terms ${\rm F}$ and ${\rm B}$ of the Lagrangian
(\ref{Lorf2}) \bea\nonumber d\chi\to D\chi =d\chi +[dh\,, \chi ]\,
, \eea where \bea\nonumber dh=H^{-1}dH ={\rm diag }
(dh_{_{\ads}}\,, dh_{_{\CP}}) = {\rm diag } (\sfrac{i}{2}\Gamma^0
dt\,,\ T_{36}\, d\phi)\,, \eea and the transformed $G$ is just the
constant matrix $G = K= {\rm diag } (K_4,K_6)$, and \bea\nonumber
dG G^{-1} = dh - K dh^t K =  {\rm diag } (i\Gamma^0 dt\,,\ (T_{36}
+K  T_{36} K) \, d\phi)=  {\rm diag } (i\Gamma^0 dt\,,\ T_{6} \,
d\phi)\, .\eea

\smallskip

Taking into account that if all bosonic fluctuations vanish
then the world-sheet metric is flat, we  conclude that the quadratic fermion Lagrangian (\ref{Lorf2}) is equal to
\bea\la{Lagf2} \L_F^{(2)}&=&{\rm str}\left[\B_{0}\pa_{0}G G^{-1} -  i\kappa {\rm F}_{0}
K\pa_1\chi^{st}{ K}^{-1}\right]\\\nonumber
&=& {\rm str}\big[ {1\ov 2}(\pa_0\chi\chi -\chi \pa_0\chi)(i \Gamma^0 + T_6)) +  i\kappa [ {i\ov 2}\Gamma^0 +T_{36}\,,  \chi  ]K\pa_1\chi^{st}{ K}\\\nonumber
&&+{1\ov 2} (({i\ov 2}\Gamma^0 +T_{36})\chi^2 +  \chi^2({i\ov 2}\Gamma^0 +T_{36})- 2\chi ({i\ov 2}\Gamma^0 +T_{36})\chi  )(i \Gamma^0 + T_6)\big]
\, ,~~~~ \eea
where we use the obvious embedding of the matrices $\Gamma^0\,,\ T_6\,,\ T_{36}$ into
$\osp$.

\smallskip

The  Lagrangian (\ref{Lagf2}) is invariant under $\kappa$-symmetry
transformations discussed in section \ref{kappasym}.  The symmetry
allows one to impose the gauge-fixing condition \bea \chi\,
T_{56}=0\,, \eea which implies that the last two columns of
$\theta$ and last two rows of $\eta$ vanish leaving only 16
physical fermion degrees of freedom.

\smallskip
Then one can easily check that due to the gauge fixing $ {\rm
str}\, (\pa_0\chi\chi -\chi \pa_0\chi)T_6 = 0\,, $ and $ {\rm
str}\,  [ T_{36}\,, \chi  ] K\pa_1\chi^{st}{ K} = 0 $. Hence, the
kinetic term in eq.(\ref{Lagf2}) becomes non-degenerate and equal
to \bea\nonumber{\rm str}\big[ {1\ov 2}(\pa_0\chi\chi -\chi
\pa_0\chi)(i \Gamma^0 + T_6)) \big] = -i\, \tr\, \eta\Gamma^0
\dot{\theta} = i\,  \tr\,  \theta^tC_4\Gamma^0 \dot{\theta} =
\tr\,  \theta^t\Gamma^3 \dot{\theta} \, , \eea and the
$\s$-derivative term is given by \bea\nonumber {\rm str}\big(
i\kappa [ {i\ov 2}\Gamma^0\,, \chi  ] K\pa_1\chi^{st}{ K}\big) =
\kappa\, \tr\, \theta^t\,\Gamma^0\,K_4\, {\theta}^{\prime} \,K_6
\, .~~~~ \eea Computing the mass term, we get \bea\nonumber &&{\rm
str}\big[{1\ov 2} (({i\ov 2}\Gamma^0 +T_{36})\chi^2 + \chi^2({i\ov
2}\Gamma^0 +T_{36})- 2\chi ({i\ov 2}\Gamma^0 +T_{36})\chi  )(i
\Gamma^0 + T_6)\big]
\\\nonumber&&
= -{1\ov 2}\, \tr\big[\theta^t C_4\theta (I - \{ T_6,T_{36}\} )
\big] \, .~~~~ \eea Finally, introducing a fermion 4 by 4 matrix
$\vartheta$ made of nonvanishing entries of $\theta$ we can write
the quadratic Lagrangian in the form (with $\kappa=1$)
\bea\la{Lf2} \L_F^{(2)} = \tr\, \big(  \vartheta^t\Gamma^3
\dot{\vartheta} + \vartheta^t\,\Gamma^0\,K_4\,
{\vartheta}^{\prime} \,K_4 - {1\ov 2}\,\vartheta^t C_4\vartheta
D_4 \big) \,, \eea where $D_4 = {\rm diag}(1,1,3,1)$ is the
restriction of  $I - \{ T_6,T_{36}\} $ to the first four entries.

\smallskip

Computing the spectrum, one finds that the Lagrangian (\ref{Lf2})
describes eight fermions with frequencies $\omega_p = \sqrt{p^2 +
{1\ov 4}}$,  four fermions with frequencies $\omega_p = - {1\ov 2}
+ \sqrt{p^2 + 1}$, and four fermions with frequencies $\omega_p =
{1\ov 2} + \sqrt{p^2 + 1}$.  It is clear from the spectrum that
the fermion Lagrangian  (\ref{Lf2}) describes eight fermions of
mass 1/2 and eight fermions of mass 1 because the constants $\pm
1/2$ in the last eight frequencies can be removed by a
time-dependent redefinition of the corresponding fermions. In
fact, the time dependence reflects the fact that some of the
fermions are still charged with respect to the U(1) subgroup that
generates the shifts of the angle variable $\phi$.

\smallskip

It is easy  to guess that  the right fermion spectrum, {\it i.e.}
the one  without any constant shifts by $\pm 1/2$ in the
frequencies, is obtained from eq.(\ref{Lf2}) by replacing $D_4$ by
the matrix $ {\rm diag}(1,1,2,2)$ which is the restriction of  $I
- T_6^2 $ to the first four entries. This replacement is just a
subtraction of the matrix ${\rm diag}(I_2,\s_3)$ from $D_4$, and
this suggests to represent the fermion $\vartheta$ in the
following  block form \bea\nonumber \vartheta =
\left(\begin{array}{cc}
\vartheta_1 & \zeta_1 \\
\vartheta_2 & \zeta_2
\end{array}\right)\,,
\eea where $\vartheta_i\,,\ \zeta_i$  are $2 \times 2$ fermion
matrices which satisfy the following hermiticity conditions \bea
\vartheta_1^\dagger = i \vartheta_2^t\s_2\,,\quad
\vartheta_2^\dagger = -i \vartheta_1^t\s_2\,,\quad \zeta_1^\dagger
= i \zeta_2^t\s_2\,,\quad \zeta_2^\dagger =-i
\zeta_1^t\s_2\,,\quad \eea where $\s_i$ are the Pauli matrices.
Computing the Lagrangian (\ref{Lf2}), one finds \bea \la{Lf3}
\L_F^{(2)} &=& \tr\, \big( 2 \vartheta_2^t\s_2\dot\vartheta_1
-\vartheta_1^t\s_2\vartheta_1' \s_2 +
\vartheta_2^t\s_2\vartheta_2' \s_2 - i
\vartheta_2^t\s_2\vartheta_1 \\\nonumber &&~~\, \,+ 2
\zeta_2^t\s_2\dot\zeta_1  -\zeta_1^t\s_2\zeta_1' \s_2 +
\zeta_2^t\s_2\zeta_2' \s_2  - 2 i \zeta_2^t\s_2\zeta_1 - i
\zeta_2^t\s_2\zeta_1 \s_3\big)\,. \eea It is now clear that the
last term in (\ref{Lf3}) can be removed by the following fermion
redefinition \bea \zeta_1\to\zeta_1 e^{i\tau\s_3/2}\,,\quad
\zeta_2\to\zeta_2 e^{-i\tau\s_3/2}\,, \eea and the first and the
second lines  (without the last term) of  eq.(\ref{Lf3})  describe
eight fermions of mass 1/2 and eight fermions of mass 1,
respectively.

\smallskip

The sum of  the quadratic bosonic and fermionic actions coincides
with the light-cone  Green-Schwarz action for type IIA
superstrings on the pp-wave background with 24 supersymmetries
constructed in  \cite{Sugiyama:2002tf,IIAlc}.

\section*{Acknowledgements}
We would like to thank Bernard de Wit for valuable discussions.
The work of G.~A. was supported in part by the RFBI grant
N05-01-00758, by the grant NSh-672.2006.1, by NWO grant 047017015
and by the INTAS contract 03-51-6346. The work of S.F. was
supported in part by the Science Foundation Ireland under Grant
No. 07/RFP/PHYF104  and by a
one-month Max-Planck-Institut f\"ur Gravitationsphysik
Albert-Einstein-Institut grant. The work of G.~A. and S.~F.~was supported in
part by the EU-RTN network {\it Constituents, Fundamental Forces
and Symmetries of the Universe} (MRTN-CT-2004-512194).

\appendix

\section{Gamma- and T-matrices}
Introduce the following matrices
\bea \label{Gmatrices}
\begin{aligned} \Gamma^0&=&{\scriptsize \left(\begin{array}{cccc}
1 & 0 & 0 & 0 \\
0 & 1 & 0 & 0 \\
0 & 0 & -1 & 0 \\
0 & 0 & 0 & -1 \\
\end{array}\right)
}\, , ~~~~~ \Gamma^1={\scriptsize \left(\begin{array}{cccc}
0 & 0 & 1 & 0 \\
0 & 0 & 0 & -1 \\
-1 & 0 & 0 & 0 \\
0 & 1 & 0 & 0 \\
\end{array}\right)
}\, , \\
\Gamma^2&=&{\scriptsize \left(\begin{array}{cccc}
0 & 0 & 0 & 1 \\
0 & 0 & 1 & 0 \\
0 & -1 & 0 & 0 \\
-1 & 0 & 0 & 0 \\
\end{array}\right)
}\, , ~~~~~ \Gamma^3={\scriptsize \left(\begin{array}{cccc}
0 & 0 & 0 & -i \\
0 & 0 & i & 0 \\
0 & i & 0 & 0 \\
-i & 0 & 0 & 0 \\
\end{array}\right)
}\, , \end{aligned}
 \eea
These matrices satisfy the Clifford algebra
$\{\Gamma^{\mu},\Gamma^{\nu}\}=2\eta^{\mu\nu}$, where
$\eta^{\mu\nu}$ is Minkowski metric with signature $(1,-1,-1,-1)$.
We also define $\Gamma^5=-i\Gamma^0\Gamma^1\Gamma^2\Gamma^3$ with
the property $(\Gamma^5)^2={\mathbb I}$.

\smallskip

The charge conjugation matrix $C_4$ obeys
$(\Gamma^{\mu})^t=-C_4\Gamma^{\mu} C_4^{-1}$ and in the present
case it can be chosen as \bea C_4=i\Gamma^0\Gamma^3={\scriptsize
\left(\begin{array}{cccc}
0 & 0 & 0 & 1 \\
0 & 0 & -1 & 0 \\
0 & 1 & 0 & 0 \\
-1 & 0 & 0 & 0 \\
\end{array}\right)
}\, . \eea

\smallskip

The Lie algebra $\alg{so}(3,2)$ is generated by the generators
$M^{ab}=-M^{ba}$ with $a,b=0,\ldots, 4$ obeying the following
relations
$$
[M^{ab},M^{cd}]=\bar{\eta}^{bc}M^{ad}-\bar{\eta}^{ac}M^{bd}-\bar{\eta}^{bd}M^{ac}+\bar{\eta}^{ad}M^{bc}\,
,
$$
where $\bar{\eta}={\rm diag}(1,-1,-1,-1,1)$. These generators have
the following representation by $4\times 4$ matrices
$M^{\mu\nu}=\sfrac{1}{4}[\Gamma^{\mu},\Gamma^{\nu}]\equiv
\Gamma^{\mu\nu}$ and $M^{\mu 4}=\frac{i}{2}\Gamma^{\mu}$. Such an
identification provides an isomorphism $\alg{so}(3,2)\sim
\alg{usp}(2,2)$ because in this representation
$(M^{ab})^{\dagger}\Gamma^0+\Gamma^0M^{ab}=0$. The matrices
$\Gamma^{\mu\nu}$ generate the Lie algebra $\alg{so}(3,1)$ and
they all commute with $\Gamma^5$. Finally, $i\Gamma^{\mu}$ span a
space of solutions to the equation $\Omega(A)=-A$ for $A$
restricted to $\alg{usp}(2,2)$.

\smallskip

The stationary subalgebra of the automorphism $\Omega$ restricted
to the $\alg{so}(6)$ component is determined by the condition
$$
[K_6,Y]=0\, , ~~~~~~Y\in \alg{so}(6).
$$
The solution to this equation can be parametrized as follows \bea
Y={\scriptsize \left(\begin{array}{rrrrrrr} 0 & y_{12} & y_{24} &
-y_{23} &
y_{26} & -y_{25} \\
-y_{12} & 0 & y_{23} & y_{24} & y_{25} & y_{26}  \\
-y_{24}  & -y_{23} & 0 & y_{34} & y_{46} & -y_{45}  \\
y_{23}  & -y_{24} & -y_{34}& 0 & y_{45} & y_{46} \\
-y_{26}  & -y_{25} & -y_{46} & -y_{45} & 0 & y_{56} \\
y_{25}  & -y_{26} & y_{45} & -y_{46} & -y_{56}& 0
\end{array}\right)}\, .
\eea This is a 9-parametric solution which describes an embedding
of the $\alg{u}(3)\subset \alg{so}(6)$.

\smallskip

The space orthogonal to $\alg{u}(3)$ in $\alg{so}(6)$ is spanned
by solutions to the following equation \bea K_6 Y=-Y
K_6\,\label{pCP}\eea and it provides a parametrization of the
coset space $\CP^3$.  The general solution to eq.(\ref{pCP}) is
six-parametric and is represented by a matrix \bea\label{Y6}
Y={\scriptsize \left(\begin{array}{rrrrrrr} 0& 0 & y_1 & y_2 &
y_3 & y_4 \\
0 & 0 & y_2 & -y_1 & y_4 & -y_3  \\
-y_1  & -y_2 & 0 & 0 & y_5 & y_6  \\
-y_2  & y_1 & 0& 0 & y_6 & -y_5 \\
-y_3  & -y_4 & -y_5 & -y_6 & 0 & 0 \\
-y_4  & y_3 & -y_6 & y_5 & 0& 0
\end{array}\right) } \equiv y_i T_i\, .
\eea Here we have introduced the six matrices $T_i$ which are Lie
algebra generators of $\alg{so}(6)$ along the $\CP^3$ directions:
\bea \label{Tmatrices}\begin{aligned}
T_1&=E_{13}-E_{31}-E_{24}+E_{42}\, ,
~~~~~~~~T_2=E_{14}-E_{41}+E_{23}-E_{32}\, , \\
T_3&=E_{15}-E_{51}-E_{26}+E_{62}\, ,
~~~~~~~~T_4=E_{16}-E_{61}+E_{25}-E_{52}\, , \\
T_5&=E_{35}-E_{53}-E_{46}+E_{64}\, ,
~~~~~~~~T_6=E_{36}-E_{63}+E_{45}-E_{54}\, ,
\end{aligned}
 \eea
where $E_{ij}$ are the standard matrix unities. The matrices $T_i$
are normalized as follows \bea {\rm tr}(T_iT_j)=-4\delta_{ij}\, .
\eea

\smallskip

The matrices $[T_i,T_j]$ commute with $K_6$ and they are
skew-symmetric\footnote{The anti-commutators $\{T_i,T_j\}$ commute
with $K_6$ as well. As the consequence, all $T_{ij}$ are
symmetric.
Not all the matrices $\{T_i,T_j\}$ are independent. In particular,
\bea\nonumber \{T_1,T_2\}=0\, ,~~~~\{T_3,T_4\}=0\, ,
~~~~\{T_5,T_6\}=0\, \eea which can be verified by a direct
calculation. From the remaining matrices $\{T_i,T_j\}$ only six
are  independent. One can choose, for instance, \bea\nonumber
\{T_1,T_4\}\, ,~~~\{T_2,T_4\}\, ,~~~ \{T_1,T_6\}\, ,~~~
\{T_2,T_6\}\, ,~~~\{T_3,T_6\}\, ,~~~ \{T_4,T_6\}\, .\nonumber
\eea}. Only 9 of them are independent and they are the generators
of $\alg{u}(3)$ inside $\alg{so}(6)$.

\smallskip

Quite remarkably, the matrix (\ref{Y6}) obeys the following
identity \bea\la{y3} Y^3=-\r^2 Y \, ,  ~~~~~~~\r^2=\sum_{i=1}^6y_i^2\,
.\eea A Lie algebra element parametrizing the coset ${\rm
AdS}_4\times \CP^3$ is therefore represented in the form \bea
A=\left(\begin{array}{cc} x_{\mu}\Gamma^{\mu} & 0 \\
0 & Y
\end{array}\right)\, , ~~~~~Y=y_iT_i\, .
\eea Thus, \bea\nonumber A^2=
\left(\begin{array}{cc} x^2{\mathbb I} ~&~ 0 \\
0 ~&~ Y^2
\end{array}\right)\, ,~~~~~A^3=
\left(\begin{array}{cc} x^2 x_{\mu}\Gamma^{\mu} ~&~ 0 \\
0 ~&~ Y^3
\end{array}\right)=\left(\begin{array}{cc} x^2 x_{\mu}\Gamma^{\mu} ~&~ 0 \\
0 ~&~ -\rho^2Y
\end{array}\right)\, ,
\eea where $x^2=x_{\mu}x_{\nu}\eta^{\mu\nu}$. Obviously, \bea {\rm
str }A^2=4x^2+4y^2\, , ~~~~{\rm str} \Sigma A^2=4x^2-4y^2\,  \eea
and as the result we find the following characteristic equation
\bea \label{rA2} A^3=\sfrac{1}{8}\, {\rm str}(\Sigma A^2)\, A +
\sfrac{1}{8}\, {\rm str}( A^2)\, \Sigma A\, ,\eea for a matrix Lie
algebra element $A$ parametrizing the space ${\rm AdS}_4\times
\CP^3$.

\section{Parametrizations of $\CP^3$}\la{apar}

An SO(6) matrix parametrizing the coset space  ${\rm SO(6)}/{\rm U}(3)$, and therefore $\CP^3$, can be obtained by exponentiating the generic  element (\ref{Y6}).
The matrix exponent can be easily computed by using the identity (\ref{y3}), and gives a
generic representative of the coset in the following form
\bea
\la{eY} g=e^Y= I+\frac{\sin\r}{\r}Y +\frac{1-\cos\r}{\r^2}Y^2\, .
\eea The formula (\ref{eY}) suggests to parametrize  $\CP^3$ by means of the spherical coordinates
\bea\la{param} y_1 + i y_2 &=& \r\, \sin \theta
\cos\frac{\alpha_1}{2} \,e^{\frac{i}{2} (\alpha_2+\alpha_3)+i \phi
} \,,\\\nonumber
   y_3 + i y_4 &=& \r\, \sin \theta \sin
 \frac{\alpha_1}{2} \,e^{-\frac{i}{2}  (\alpha_2-\alpha_3)+i \phi } \,,\\\nonumber
   y_5 + i y_6 &=& \r\,  \cos \theta \, e^{i \phi } \,.
\eea This provides an explicit parametrization of $\CP^3$ which
can be used to find the Fubini-Study metric on $\CP^3$. To this
end, by noting that $K_6g=g^{-1}K_6$ we first compute
\bea\nonumber -2A^{(2)} = g^{-1}dg + K_6g^{-1}dgK_6 = g^{-1}dg +
dgg^{-1}\,. \eea Then, the Fubini-Study metric on $\CP^3$ is given
by the following formula
 \bea ds^2_{_{\CP^3}} &=&
-{1\ov 4} \tr \left(A^{(2)}\right)^2  = {1\ov 8} \tr\left( dg
dg^{t} - g^{t}g^tdgdg\right) = {1\ov 16} \tr\left( d(g^2)
d(g^{2})^t\right)\\\nonumber &=& d\r^2 + {1\ov4}\sin ^22\r \left(
d\phi+\frac{1}{2}\sin^2\theta\left(d\alpha_3+d\alpha_2\cos
\alpha_1\right) \right)^2 + \sin^2\r\, ds^2_{_{\CP^2}}\,, \eea
where \bea\nonumber ds^2_{_{\CP^2}} =  d\theta^2 + {1\ov
4}\sin^2\theta\left( d\a_1^2 + \sin^2\a_1 d\a_2^2 +
\cos^2\theta\left( d\a_3 + \cos\a_1 \, d\a_2\right)^2\right) \eea
is  the Fubini-Study metric on $\CP^2$.

\smallskip

Note also that $\CP^3$ can be also parametrized by means of the
following matrices
\bea G = g K_6 g^t = g^2 K_6\,,\quad
G^t=-G\,,\quad GG^t=I\,, \eea and therefore $\CP^3$ can be
identified with the intersection of skew-symmetric and orthogonal
matrices. In terms of the matrix $G$ the Fubini-Study metric on
$\CP^3$ is given by the following simple formula
 \bea ds^2_{_{\CP^3}} ={1\ov 16} \tr\, dG
dG^t\,. \eea

\smallskip

It is well known that the Fubini-Study metric on  $\CP^3$ can be also written in the form
\bea\la{metricw} ds^2_{_{\CP^3}} = {d {\bar w}_i d w_i\ov 1 +
 |w|^2}\  -\  { d {\bar w}_i w_i  {\bar w}_jd w_j \ov (1
+  |w|^2)^2}\,,\qquad  |w|^2= {\bar w}_k  w_k\,. \eea The three
complex inhomogeneous coordinates $w_i$ are related to the six real coordinates
$y_i$ as follows \bea \nonumber
&&\hspace{-0.35cm}\sin\r = {|w|\ov\sqrt{1 + |w|^2}}\,,\  \cos\r = {1\ov\sqrt{1 + |w|^2}}\,, \quad \sin 2\r = {2|w|\ov 1 + |w|^2}\,,\  1 - \cos2 \r = {2|w|^2\ov 1 + |w|^2}\,, \\
&&{|w|\ov\r}(y_1+iy_2) = w_1\,,\quad {|w|\ov\r}(y_3+iy_4) =
w_2\,,\quad {|w|\ov\r}(y_5+iy_6) = w_3\,. \eea
Then, the coset
representative $g$ takes the form
 \bea\la{gcos} g = I + {1\ov\sqrt{1 +
|w|^2}} ( W + {\overline W})  + { \sqrt{1 + |w|^2} - 1\ov |w|^2
\sqrt{1 + |w|^2}} ( W {\overline W} +   {\overline W} W)\,, \eea
where \bea\label{Wp}\begin{aligned} &W = w_i {\cal T}_i\,,  \quad
~{\cal T}_1 = {1\ov 2}(T_1 - i T_2) \,,  \quad ~{\cal T}_2 = {1\ov
2}(T_3 - i T_4) \,, \quad {\cal T}_3 = {1\ov 2}(T_5 - i T_6) \,,\\
&W = {\bar w}_i {\overline {\cal T}}_i\,,  \quad {\overline {\cal
T}}_1 = {1\ov 2}(T_1 + i T_2) \,,  \quad {\overline {\cal T}}_2 =
{1\ov 2}(T_3 + i T_4) \,,  \quad {\overline {\cal T}}_3 = {1\ov
2}(T_5 + i T_6) \,,~~~~~~~~
\end{aligned}\eea and we took into account that $W^2 =0$ for any set of
$w_i$. Computing $g^2$ we get the following simple formula \bea
g^2 &=& I + {2\ov 1 + |w|^2} ( W + {\overline W})  + {2\ov 1 +
|w|^2} ( W {\overline W} + {\overline W} W)\\\nonumber & =&  -I \,
{1- |w|^2 \ov 1+ |w|^2 }\ +\   {2(I + W)(I+{\overline W})\ov 1 +
|w|^2} \,, \eea which can be used to find $G$ and verify
(\ref{metricw}).

\smallskip

The $\CP^3$ metric can be  written
in terms of the four homogeneous coordinates $z_a$ \bea\la{metricz}
ds^2_{_{\CP^3}} = {d {\bar z}_a d z_a\ov {\bar z}_c  z_c}\  -\  {
d {\bar z}_a z_a  {\bar z}_b d z_b \ov ( {\bar z}_c  z_c)^2}\,,
\eea which is the standard form of the Fubini-Study metric.
Inhomogeneous coordinates $w_i$ are related to $z_a$ as follows \bea w_i = {z_i\ov
z_4}\,, \eea and the metric (\ref{metricz}) obviously reduces to
(\ref{metricw})  if $z_4=1$.

\smallskip

It is clear from (\ref{metricw}) that there are 3 commuting
isometry directions corresponding to multiplying $w_i$ by a phase
$w_i\to e^{i\a}w_i$.


\end{document}